# From itinerant ferromagnetism to insulating antiferromagnetism: A magnetic and transport study of single crystal SrRu$_{1-x}$Mn$_x$O$_3$ (0≤ x<0.60)

G. Cao, S. Chikara, X.N. Lin, E. Elhami and V. Durairaj
Department of Physics and Astronomy
University of Kentucky, Lexington, KY40506

P. Schlottmann
Department of Physics and National High Magnetic Field Laboratory
Florida State University, Tallahassee, FL32306

We report results of a magnetic and transport study of SrRu$_{1-x}$Mn$_x$O$_3$ (0≤ x<0.60), i.e., Mn doped SrRuO$_3$. The Mn doping drives the system from the itinerant ferromagnetic state (T$_C$=165 K for x=0) through a quantum critical point at x$_c$=0.39 to an insulating antiferromagnetic state. The onset of antiferromagnetism is abrupt with a Néel temperature increasing from 205 K for x=0.44 to 250 K for x=0.59. Accompanying this quantum phase transition is a drastic change in resistivity by as much as 8 orders of magnitude as a function of x at low temperatures. The critical composition x$_c$=0.39 sharply separates the two distinct ground states, namely the ferromagnetic metal from the antiferromagnetic insulator.

**Introduction**

In numerous strongly correlated electron systems different degrees of freedom, such as the spin, orbitals and lattice deformations are inextricably coupled, usually by Coulomb interactions and the specifics of the crystal structure, but also due to alloying. Such systems are often characterized by competing ground states susceptible to external perturbations such as magnetic field, pressure or chemical doping. Tuning the external parameters may lead to a quantum critical point and stabilize novel ground states with exotic properties. This point is well illustrated in the cuprate superconductors [1].

The layered ruthenates, i.e., the Ruddlesden-Popper (RP) series $Ca_{n+1}Ru_nO_{3n+1}$ and $Sr_{n+1}Ru_nO_{3n+1}$ (n=number of Ru-O layeres/unit cell), are a class of correlated electron materials showing a rich variety of properties. The major characteristic of these 4d-electron based transition metal oxides is the more extended d-orbitals of the Ru-ion as compared to those of 3d ions, which greatly enhances the transition-metal-oxygen or *p-d* hybridization. The physical properties of the ruthenates are critically linked to n, the number of Ru-O layers per unit cell, and to the cation (Ca or Sr), which lead to different ground states and inter- and intra-layer magnetic couplings. The Ru-ions are surrounded by O-ions forming octahedra. The deformations and relative orientations of these corner-shared octahedra crucially determine the crystalline field splitting, the band structure and hence the magnetic and transport properties. As a result, the $Sr_{n+1}Ru_nO_{3n+1}$ are metallic and tend to be ferromagnetic with $Sr_2RuO_4$ (n=1) being an exception, whereas the isoelectronic $Ca_{n+1}Ru_nO_{3n+1}$ are all at the verge of a metal-nonmetal transition and prone to antiferromagnetism. The Curie temperature, $T_C$, for the series $Sr_{n+1}Ru_nO_{3n+1}$ increases with n, whereas the Néel temperature, $T_N$, for the $Ca_{n+1}Ru_nO_{3n+1}$ series decreases with increasing n [2-12]. No such behavior has been observed in other transition metal RP systems. Controlling the orientation of the octahedra by changing the chemical composition therefore opens a unique opportunity to systematically tune physical properties in these materials.

The perovskite $SrRuO_3$ is the by far the most studied compound in the $Sr_{n+1}Ru_nO_{3n+1}$ series (see for example [13-29]). $SrRuO_3$ is an itinerant ferromagnet with $T_C$=165 K and a saturation moment of 1.10 $\mu_B$/Ru aligned within the basal plane [19]. The crystal field effect in $Ru^{4+}$ ($4d^4$) ions is so large that the Hund's rules partially break down, yielding a low spin state with S=1 ($^3T_{1g}$). $SrRuO_3$ features robust Fermi-liquid behavior [17,19,28] with an enhanced effective mass [19,29] and quantum oscillations [29] at low temperatures, and anomalous transport behavior at high temperatures evidenced by a linear temperature dependence of the resistivity up to 900 K that violates the Mott-Ioffe-Regel limit [28]. It is believed that the hybridization between O $2p$ states and Ru $4d$ states and the interplay with the Mott-Hubbard limit account for the unusual magnetic and transport behavior [20,25]. Our earlier study on single crystal $Sr_{1-x}Ca_xRuO_3$ indicates that the magnetic coupling is highly sensitive to perturbations in the Ru-O-Ru bond length and angle caused by substituting Sr with the isoelectronic smaller Ca-ion. This gives rise to a rotation of the $RuO_6$ octahedron and thus a subtle change in the electron hopping between octahedra, yielding a state that is less favorable for ferromagnetism. Consequently, $T_c$ decreases monotonically with Ca concentration and vanishes for x<0.8 [19]. A similar interplay between lattice and electronic properties also leads to an antiferromagnetic insulating ground state with $T_N$=26 K in the double-perovskite $Sr_2YRuO_6$ that is derived from the perovskite $SrRuO_3$ by replacing every second Ru by Y [30].

A very different interplay between spin, orbital and lattice degrees of freedom is found in the manganite alloys $La_{1-x}Ca_xMnO_3$ and $La_{1-x}Sr_xMnO_3$ [31,32]. The Mn ions exist in their trivalent and tetravalent states. Here the Hund's rule energy aligning all spins of the 3d-electrons is relatively larger than the crystalline fields. Hence, the total spin of $Mn^{3+}$ is S=2 ($^5E_g$), while that of $Mn^{4+}$ is S=3/2 ($^4A_{2g}$). The latter state consist of the three $t_{2g}$ being singly occupied, while for $Mn^{3+}$ there is in addition one $e_g$ electron. The $t_{2g}$ electrons are localized and the interaction among the $t_{2g}$ on neighboring Mn ions is via antiferromagnetic superexchange mediated by the O bonds. The $e_g$ electrons can hop between sites and this process induces the ferromagnetic double-exchange interaction. The interplay between the double-exchange and the superexchange in conjunction with the Jahn-Teller effect (there are two $e_g$ orbitals) then leads to several ferromagnetic, antiferromagnetic and canted spin phases, the colossal magneto-resistance, orbital order, metal-insulator transitions and phase separation. The different size of $Ca^{2+}$ and $Sr^{2+}$ also plays a role in the manganites, which manifests itself, e.g., in a much larger phase separation in the Ca-compound than for the Sr-alloy.

In particular, the perovskite $SrMnO_3$ has only $Mn^{4+}$ ions, is an insulator, and has cubic symmetry with the S=3/2 $t_{2g}$-spins ordering antiferromagnetically in the G-phase (alternating up- and down-spins on a simple cubic lattice) [33]. This is possibly the simplest scenario for the manganites, because there are no $e_g$ electrons involved in the ground state. In this paper we study the system $SrRu_{1-x}Mn_xO_3$ (x<0.60), i.e. the alloy of $SrRuO_3$ with $SrMnO_3$. The different crystalline structure of the end-compound impedes to extend the range of x to larger values. This work reveals a rich phase diagram where Mn doping effectively drives the system from the itinerant ferromagnetic state through a quantum critical point at $x_c=0.39$ to an insulating antiferromagnetic state. This transition to the antiferromagnetic state occurs abruptly (but continuously) with the disappearance of the ferromagnetism. The Néel temperature increases from essentially zero at $x_c$ to 205 K for x=0.44 to 250 K for x=0.59. Accompanying this magnetic phase transition is a dramatic change in resistivity at low temperatures by as much as 8 orders of magnitude (measured at 7 K) for the entire Mn doping range of $0 \leq x < 0.60$. A Mott-type transition and quantum critical point at the composition $x_c=0.39$ sharply divides the two regimes, namely the ferromagnetic metal from the antiferromagnetic insulator. This behavior is in contrast to that seen in $Sr_{1-x}Ca_xRuO_3$ where the magnetic phase transition occurs between the ferromagnetic and paramagnetic state without an occurrence of a metal-insulator transition [19].

It is noted that some of the results presented here are different from those of a study on polycrystalline $SrRu_{1-x}Mn_xO_3$ ($x \leq 0.50$) reported earlier by Sahu et al. [34], where the Curie temperature $T_C$=165 K for x=0 remained essentially unchanged for the entire Mn doping range of $0 \leq x \leq 0.50$ (see Fig. 2a of Ref. 34). The discrepancy may, at least partially, result from the difference between single crystals and polycrystalline samples. The $SrRuO_3$ phase is notoriously known to be unavoidable in polycrystalline oxides whenever Sr and Ru ions are simultaneously involved. The unwanted ferromagnetic phase, even a very small amount of it, could overshadow changes in physical properties due to Mn doping.

**Experimental**

The single crystals of the entire series of SrRu$_{1-x}$Mn$_x$O$_3$ were grown using both floating zone and flux techniques. All crystals studied were characterized by single crystal or powder x-ray diffraction, EDS and TEM. No impurities and intergrowth were found. The fact that quantum oscillations are observed in SrRuO$_3$ [29] confirms the high quality of the crystals studied. The Dingle temperature $T_D$ estimated from the quantum oscillations, a measure of scattering rate, is in a range of $T_D$ = 2.0 K [29], comparable to those of good organic metals, whose $T_D$ varies from 0.5 to 3.5 K. The magnetization was measured using the Quantum Design MPMS LX 7T magnetometer. The resistivity was obtained using the standard four-lead technique and a function of transport measurements added to the magnetometer. For each composition, a few crystals were measured and the data was averaged in order to reduce errors that could be generated by uncertain lead geometry.

**Results**

Shown in Fig.1 are the lattice parameters for the a-, b- (left scale) and c-axis (right scale) as a function of Mn concentration, x, ranging from 0 to 0.59 for SrRu$_{1-x}$Mn$_x$O$_3$. The lattice parameters are determined using x-ray diffraction data on powdered crystals. For x=0 (SrRuO$_3$), the lattice parameters are in good agreement with those reported earlier [35]. The orthorhombic symmetry is retained as a function of x, which is consistent with results reported previously [34]. Within the error of the measurement, the lattice parameters systematically decrease with x, since the ionic radius of Mn$^{4+}$ (0.53 Å) is smaller than that of Ru$^{4+}$ (0.62 Å). The changes in the lattice parameters result in a shrinkage of the unit-cell volume by about 4%. The decrease in the a- and c-axis is more rapid near x=0.39. As seen below, this distinct behavior is accompanied by drastic changes in the electronic and magnetic structure.

Shown in Fig. 2a is the temperature dependence of the basal plane magnetic susceptibility, $\chi$, defined as M/H, for representative compositions. A similar behavior is seen for the c-axis magnetic susceptibility and therefore not shown. The ferromagnetic order decreases with x and disappears at $x_c$=0.39. The Curie temperature $T_C$ is effectively suppressed from 165 K for x=0 to lower temperatures with increasing x and vanishes for $x_c$=0.39 (the curve of $\chi$ for $x_c$=0.39 is enlarged by a factor of 20 for clarity). $\chi$ for the critical composition $x_c$=0.39, where the lattice parameters for the a- and c-axis decrease significantly (see Fig.1), displays a rather weak temperature-dependence without any sign of long-range ordering (this point is to be further discussed below) and spin glass behavior (as no hysteresis behavior is discerned). For x>$x_c$ antiferromagnetic long-range order emerges abruptly as x increases. Hence, the disappearance of the ferromagnetic state is immediately followed by the appearance of a strong antiferromagnetic state, which for x=0.44 has already a Néel temperature $T_N$=205 K. The temperature dependence of $\chi$ in the vicinity of $T_N$ is also unusual (note that $\chi$ for x=0.44 is plotted by reducing its magnitude by a factor of 25 for comparison with that of x=0.51 and 0.59). While $T_N$ further rises and eventually reaches 250 K for x=0.59, the magnitude of $\chi$ decreases with x. The change in the magnetic ground state is also reflected in the basal plane isothermal magnetization, M, shown in Fig. 2b, where M as a function magnetic field B for T=5 K is plotted for a few representative compositions. M for x≤ 0.28 displays field dependence typical of a ferromagnet, whereas the linear field dependence of M for x=0.39 and 0.59 is more characteristic of an antiferromagnet.

The data in Fig. 2a was fitted to a Curie-Weiss law $\chi=\chi_o+C/(T-\theta)$, where $\chi_o$ is a temperature-independent susceptibility, C is the Curie constant, and $\theta$ the Curie-Weiss temperature, for 200<T<350 K and 0<x<0.39. The fitting range for x=0.44 is 280-360 K since $T_N$=205 K, and no fitting was attempted for x=0.51 and x=0.59 as the temperature range is not wide enough to obtain valid parameters, again because of the high Néel temperature. As expected, the Curie-Weiss temperature $\theta$ decreases from +167 K for x=0 to -2 K for x=0.39 and finally to -242 K for x=0.44. The change in sign is associated with the change from ferromagnetic to antiferromagnetic exchange coupling. The effective moment estimated from the Curie constant C varies monotonically from 2.72 $\mu_B$ for x=0 to 3.41 $\mu_B$ for x=0.44. These values are slightly smaller but close to those anticipated for tetravalent Ru and Mn ions, i.e. S=1 for $Ru^{4+}$ and S=3/2 for $Mn^{4+}$, respectively. Note that the effective magnetic moments for $Ru^{5+}$ (S=3/2) and $Mn^{3+}$ (S=2) are considerably larger, so that the Curie constant is only consistent with tetravalent Ru and Mn ions for all compositions.

As shown in Fig. 3, the temperature independent susceptibility $\chi_o$ only depends weakly on x for x<0.24, but increases significantly at x=0.28 and peaks at $x_c$=0.39 where the ferromagnetism disappears. For x=0.51 and 0.59 the temperature range of the fit is too small and a valid value of $\chi_o$ cannot be determined. This temperature independent susceptibility is usually associated with a Pauli susceptibility and a measure of the density of states at the Fermi level, $N(\varepsilon)$, i.e., $\chi_o \sim N(\varepsilon)$. The rapid increase of $\chi_o$ near x=0.39 can then be attributed to an increase in the density of the states. At the same time, the ordered moment, $M_s$, obtained by extrapolating M to zero-field, B=0, decreases from 1.10 $\mu_B$ for x=0 to nearly zero for x=0.39. As illustrated in Fig.3, the variations of both $\chi_o$ and $M_s$ (right scale) are larger in the vicinity of x=0.39, suggesting an intimate correlation between the density of states and the disappearance of the ferromagnetism.

The low temperature resistivity also undergoes dramatic changes with x. As seen in Fig. 4a, $\rho$, varies by 8 orders of magnitude from x=0 to x=0.59 at 7 K. Shown in Fig. 4a is the resistivity in the basal plane, $\rho_{ab}$, on a logarithmic scale as a function of temperature for x=0, 0.28, 0.39 and 0.59. $\rho_{ab}$ rapidly increases with increasing x, and x=0.28 (or slightly larger) appears to be the dividing line between metallic and insulating behavior. For x=0 the resistivity is the expected one for a metal, while for x=0.59 the behavior is clearly insulating.

For x=0, there is a sharp break in the slope of $\rho$ at $T_c$ = 165 K (it is seen more clearly on a linear scale for $\rho$ [28]), which according to the Fisher-Langer theory is the consequence of scattering off short-range spin fluctuations in the neighborhood of $T_c$ [27]. The residual resistivity ratio [RRR=$\rho$(300K)/$\rho$(2K)] for x=0 is 120 and the residual resistivity $\rho_o$ is 2 $\mu\Omega$ cm for the basal plane [28]. Recently, de Haas van Alphen oscillations have been observed in this system with frequencies ranging from 100 to 11,000 T [29]. Although $\rho$ for T>Tc increases almost linearly with temperature up to 900 K and violates the Mott-Ioffe-Regel limit [28], $SrRuO_3$ behaves like a Fermi liquid for temperatures below 38 K, as $\rho$ can be described by $\rho=\rho_o+AT^2$ with A being 9.3×10$^{-9}$ $\Omega$ cm K$^{-2}$ [28]. The coefficient A is proportional to the square of the effective mass, m*, i.e. A~m*$^2$. The value of A is comparable to those of other correlated electron systems. The Fermi liquid behavior at low temperatures survives with increasing x up to x=0.28 for T≤ 24K with an enhanced A of $2.6\times10^{-8}$ $\Omega$ cm $K^{-2}$, suggesting an increase of the correlation effects with x.

At x=0.39 the ferromagnetism is already suppressed and antiferromagnetism has not yet developed as shown in Fig. 4b. For the temperature range $40\leq T\leq100$ K the resistivity can be described by $\rho=\rho_o+AT^2$ with $A=3.0\times10^{-8}$ $\Omega$ cm $K^{-2}$, i.e. the parameter A is further enhanced. The increase in A is consistent with the increase in $\chi_o$ shown in Fig.3, suggesting that the correlation effects are important in driving the metal-insulator transition. Below 20 K, with decreasing temperature, $\rho$ rises rapidly by a factor of 30 but follows no activation law or any other power law. The rapid uprising in $\rho$ is suggestive of a gap opening, e.g. due to a critical value U/W for a Mott-like transition in the Hubbard model (here W is the d-band width and U the Coulomb repulsion). This behavior is also mirrored in $\chi$. As seen in Fig. 4b, the temperature dependence of $\chi$ (right scale) at low temperatures is strikingly similar to that of $Sr_3Ru_2O_7$, which is known to have a field-tuned quantum critical point [10], and Pd, which is close to a Stoner ferromagnetic instability [36].

A further increase of x drives the system into a much more insulating state. $\rho$ for x=0.59 obeys no activation law but follows the variable range hopping behavior, $\rho\sim\exp(T_o/T)^n$, as shown in Fig. 5. Acceptable fits can be obtained for n=1/3 with $T_0=21622$ K and n=1/2 with $T_0=1220$ K, but n is probably closer to 1/3. This corresponds to two-dimensional hopping without interactions between the electrons, or the usual three-dimensional variable ranging hopping behavior with n=1/4 with some influence from long-range Coulomb repulsions, which ideally reduce n to 1/2 [37, 38]. Note also that variable range hopping conduction, which normally is expected to take place only at very low temperatures, is found over nearly the entire temperature range of 2K<T<300K for x=0.59.

**Discussion and Conclusions**

The magnetic and electronic phase diagram as a function of Mn content x is summarized in Fig. 6. It is clear that the substitution of itinerant Ru $t_{2g}$ electrons by localized Mn $t_{2g}$ electrons strongly affects the properties. As illustrated in the phase diagram, the ferromagnetism-antiferromagnestim transition is accompanied by the metal-insulator transition. Separating the two phases is the critical composition $x_c$=0.39, which corresponds to a quantum critical point [1]. Note that the transition from ferromagnetism to antiferromagnetism is not of first order, since both order parameters vanish at $x_c$. From the resistivity data it is difficult to exactly determine the concentration x of the metal insulator transition.

In SrRuO$_3$ some of the 4d $t_{2g}$-orbitals are itinerant due to self-doping by the O 2p-electrons (an effect also known as valence admixture) and the system is metallic. Transport and magnetic properties strongly depend on the relative orientation of the corner-shared octahedra. There is a strong coupling of lattice, charge, orbital and spin degrees of freedom. In SrRu$_{1-x}$Mn$_x$O$_3$ the substitution of Ru by Mn eliminates one of the $t_{2g}$ electrons and hence the itinerant character of the d-electrons. Since the Mn $t_{2g}$ levels are all occupied with one electron, the only possibility of hopping would be to temporarily fill one of the $e_g$ levels. This process is, however, energetically unfavorable. Hence, as a consequence of the large crystalline field splitting in the MnO$_6$ octahedra, the Mn sites interrupt the dynamics of the 4d $t_{2g}$-electrons, and the metallic and ferromagnetic character of the Ru end-compound gradually disappears with increasing x.

Tentatively, the metal-insulator transition could be associated with a site percolation of nearest neighbor Ru-Ru bonds. For a simple cubic lattice the percolation threshold is roughly at a Ru concentration (1-x) of 30.7%, i.e. x=0.693 [39]. This is much larger than the observed value of $x_c$=0.39. However, the SrRuO$_3$ compound is not cubic, but layered. Reducing the dimensionality reduces the connectivity and hence increases the percolation threshold (1-x). For the square lattice the critical concentration of site percolation of nearest neighbor bonds is 0.590±0.010 [39], corresponding to $x_c$=0.41, which is very close to the measured $x_c$ for SrRu$_{1-x}$Mn$_x$O$_3$. This agreement, however, is likely to be accidental since the compound is not strictly two-dimensional. The disruption of Ru connectivity also affects the orientation of the RuO$_6$ octahedra (tilting angle), which to a great extent determines the properties of the ruthenates.

Decreasing the connectivity of the Ru bond network also reduces the ferromagnetic coupling, which is probably of the double-exchange type, i.e. induced by the hopping of the 4d $t_{2g}$ electron in the highest energy level (note that hopping requires self-doping through the oxygen orbitals). Experimentally it is not clearly determined if the long-range ferromagnetism collapses simultaneously with the conductivity. The reduced connectivity also tends to localize the electrons and hence it leads to an increased density of states at the Fermi level. For larger x the antiferromagnetic superexchange between Mn ions, mediated by the O bonds, prevails and the alloy becomes antiferromagnetic.

Like $Mn^{4+}$, the $Fe^{4+}$ ($3d^4$) ion of the perovskite $SrFeO_3$ is in a high spin state. However, unlike Mn doping, Fe substitution, i.e. $SrRu_{1-x}Fe_xO_3$, leads to the occurrence of a paramagnetic state following a disappearance of the ferromagnetism for x>0.35. Moreover, this magnetic phase transition is not accompanied by a metal-insulator transition [40]. $Fe^{4+}$ is most likely iso-electronic to $Mn^{3+}$, i.e., three of 3d electrons of $Fe^{4+}$ occupy one $t_{2g}$ orbital each and the fourth one is in an $e_g$-orbital. Hence, the main difference between $Mn^{4+}$ and $Fe^{4+}$ is the occupation of the $e_g$-orbital. In $SrRu_{1-x}Fe_xO_3$ this $e_g$-electron may become itinerant and hybridize with the Ru $t_{2g}$ electrons.

**Acknowledgements**

This work was supported by National Science Foundation under grant No. DMR-0240813. Work at the National High Magnetic Field Laboratory was supported through NSF Cooperative Agreement DMR-0084173 and the State of Florida. P.S. acknowledges the support by NSF (grant No. DMR01-05431) and DOE (grant No. DE-FG02-98ER45707).

31. J.M.D. Coey, M. Viret, and S. von Martech, Adv. in Phys. **48**, 167 (1999)
32. T.A. Kaplan and S.D. Mahanti (eds.), *Physics of manganites*. Kluwer Academic/Plenum Publishers, New York (1999)
33. T.Takeda and S. Ohara, J. Phys. Soc. Jpn. **37**, 275 (1974)
34. Ranjan K. Sahu, Z. Hu, Manju L. Rao, et al. Phys. Rev. B **66**, 144415 (2002)
35. For example, C.W. Jones *et al*., Acta Crystallogr., Sec. C, **45**, 365 (1989)
36. W. Gerhardt, F. Razavi, J.S. Schilling, D. Huser and J.A. Mydosh, Phys. Rev. B **24**, 6744 (1981).
37. A.L. Efros and M. Pollak (eds.), *Electron-electron interactions in disordered systems*. Elsevier, Amsterdam (1986)
38. A.L. Efros and B.I. Schlovksii, J. Phys. C: Solid State Phys. **8**, L49 (1975).
39. J.W. Essam, in *Phase Transitions and Critical Phenomena*, edited by C. Domb and M.S. Green, (Academic Press, London, 1972), vol. 2, p. 197.
40. S. Chikara et al., to be published elsewhere.

Captions:

**Fig.1.** Lattice parameters for the a-, b- (left scale) and c-axis (right scale), for $SrRu_{1-x}Mn_xO_3$ as a function of Mn concentration, x, which ranges from 0 to 0.59.

**Fig.2.** (a) Temperature dependence of basal plane magnetic susceptibility $\chi$ for representative compositions (left scale for $x \leq 0.39$, and right scale for the remaining x). Note that $\chi$ for x=0.39 is enlarged by a factor of 20 for clarity; and that $\chi$ for x=0.44 is reduced by a factor of 25 for comparison with that for x=0.51 and x=0.59; (b) Basal plane isothermal magnetization M as a function of magnetic field B at T=5 K for a few representative x.

**Fig.3.** Temperature independent magnetic susceptibility $\chi_o$ and the saturation moment Ms (right scale) as a function of x. Note that $\chi_o$ for x=0.51 and 0.59 cannot be obtained because the temperature is too narrow for a valid fitting.

**Fig.4.** (a) Basal plane resistivity $\rho_{ab}$ on a logarithmic scale vs. temperature for x=0, 0.28, 0.39, and 0.59; (b) $\rho$ and $\chi$ (right scale) for $x_c$=0.39 as a function of temperature for $2 \leq T \leq 120$ K for comparison.

**Fig.5.** Variable range hopping plot for the resistivity. The logarithm of the resistivity for x=0.59 is presented against $T^{-1/3}$ and $T^{-1/2}$ (lower and upper horizontal axis, respectively) for a wide temperature range (2<T<300 K).

**Fig.6.** T-x phase diagram illustrating the crossover from itinerant ferromagnetism to the insulating antiferromagnetism. There is a quantum critical point (QCP) at $x_c$=0.39 separating the two quantum ground states. F-M stands for ferromagnetic metal and AF-I antiferromagnetic insulator.

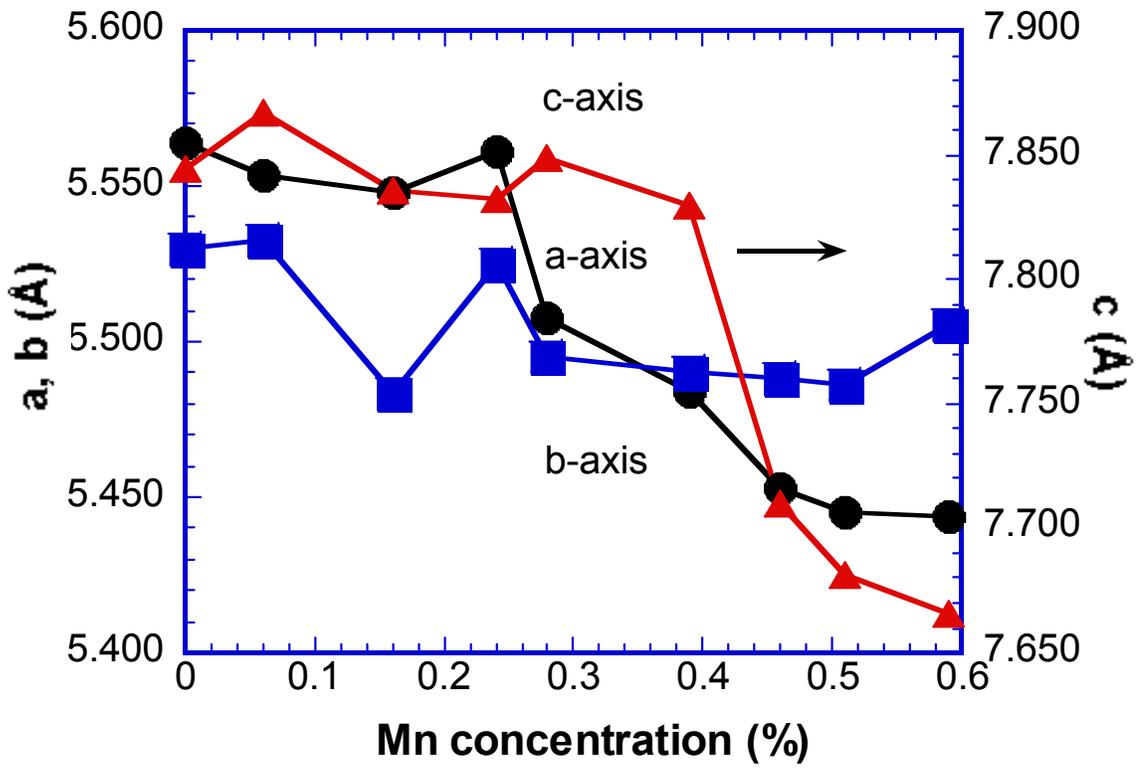

Fig.1

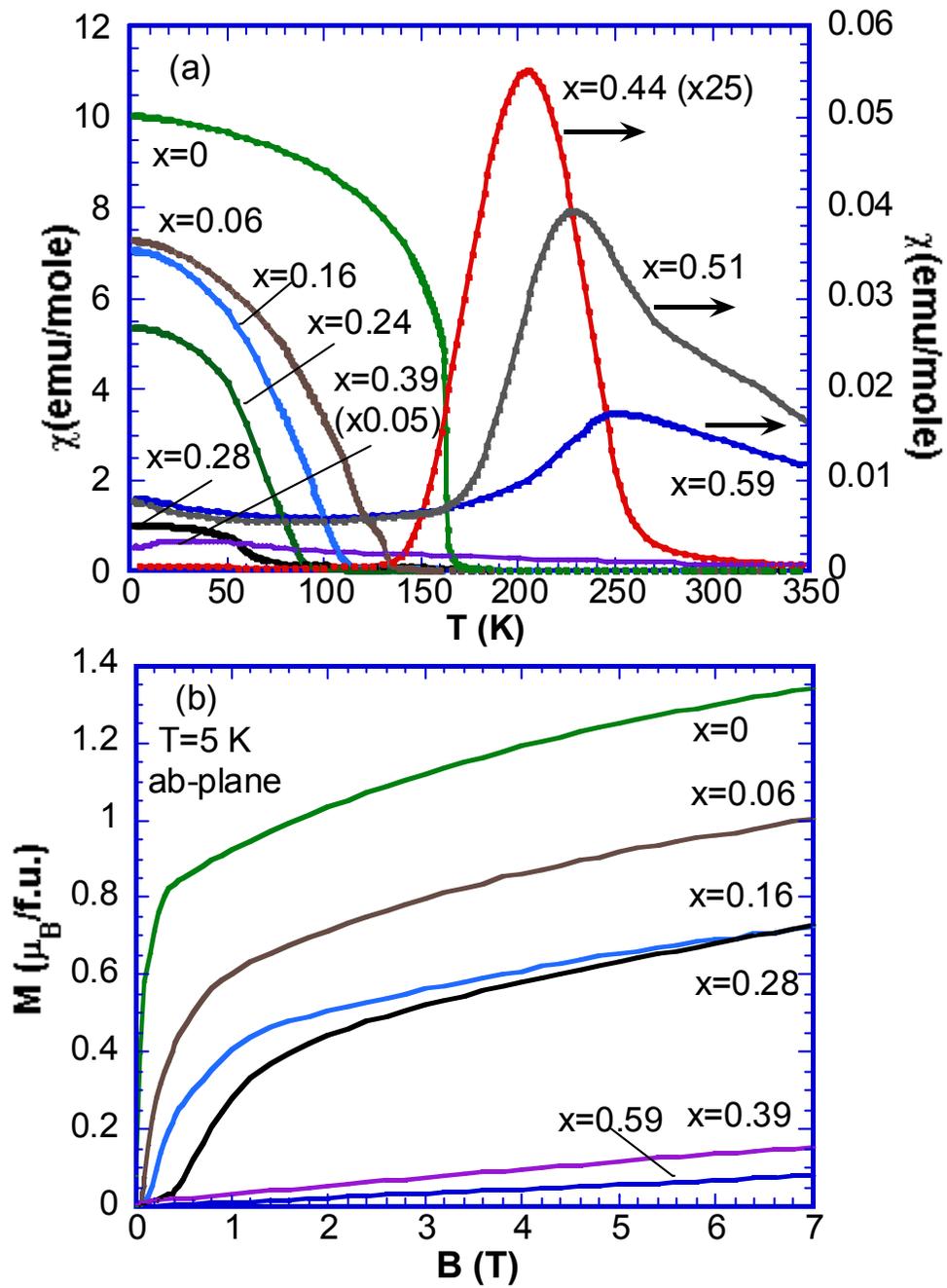

Fig.2

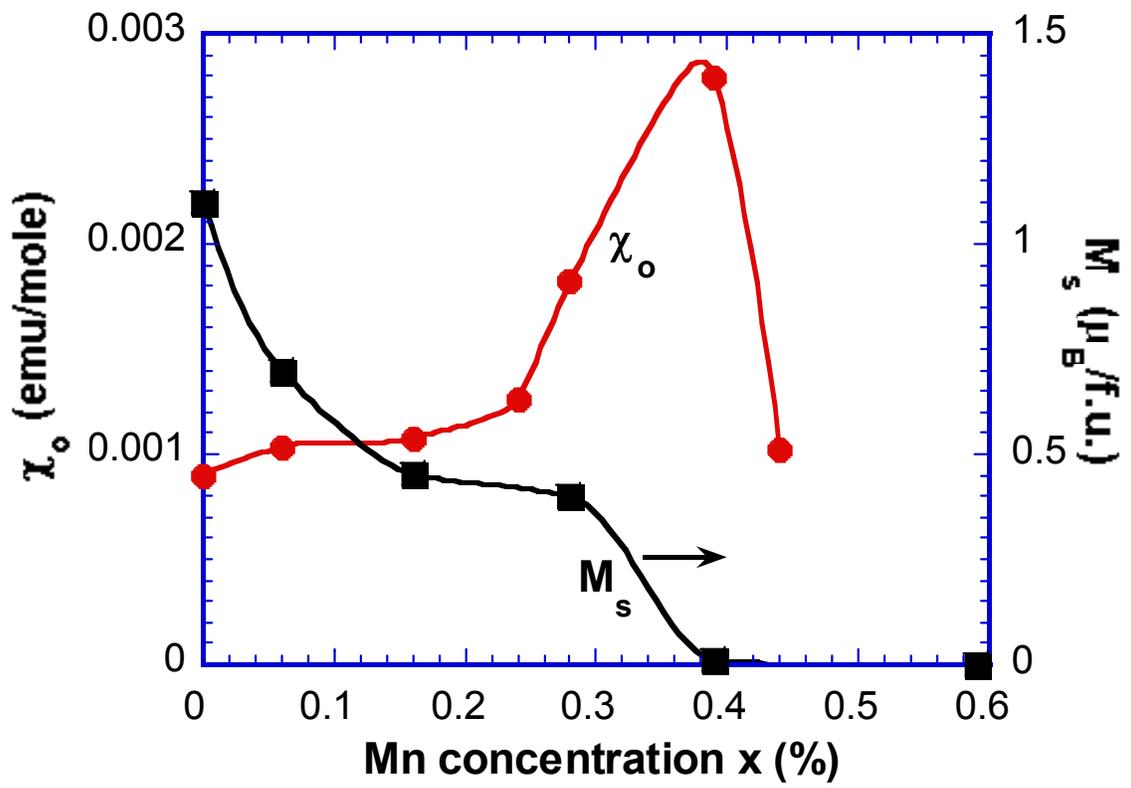

Fig.3

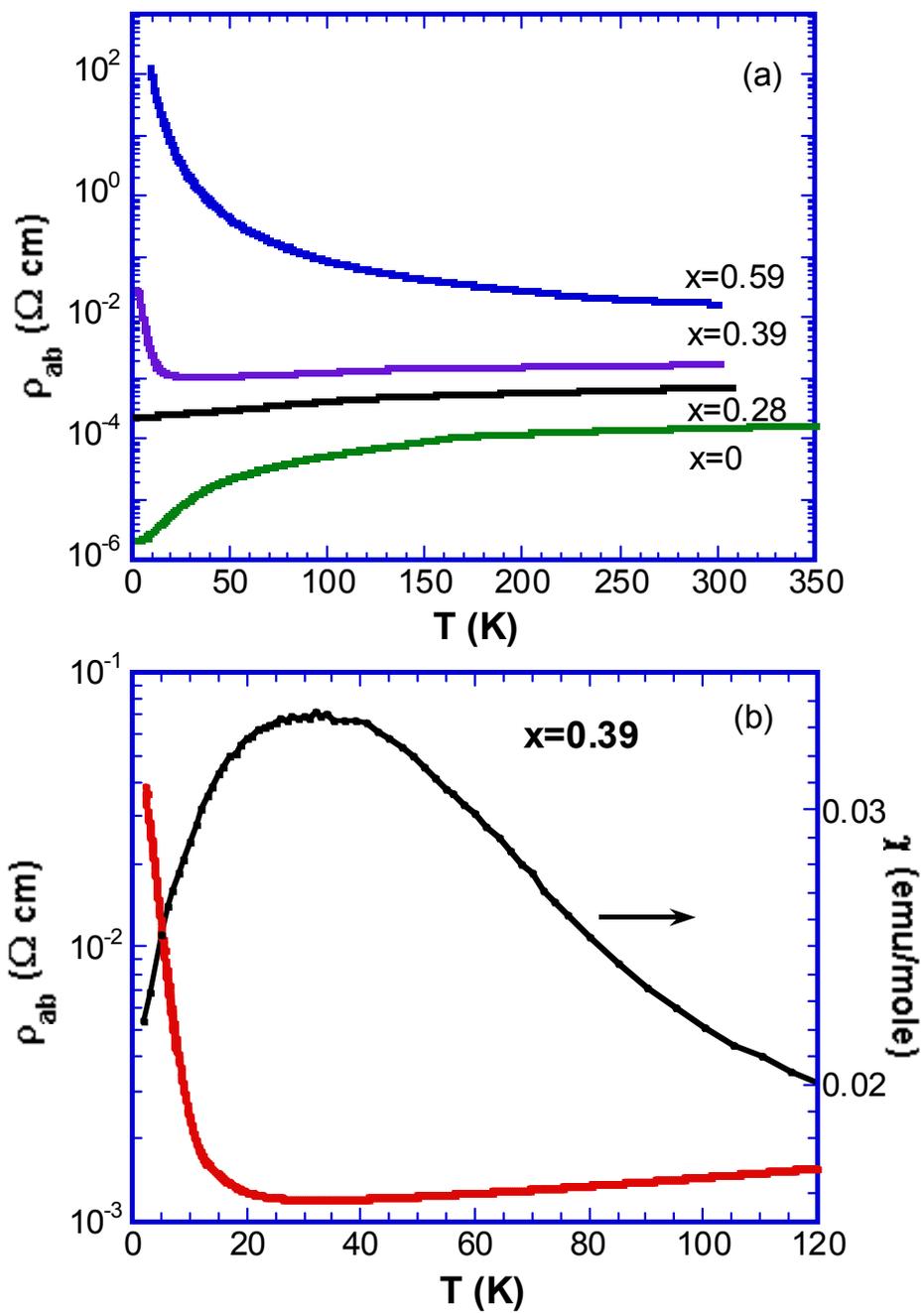

Fig.4

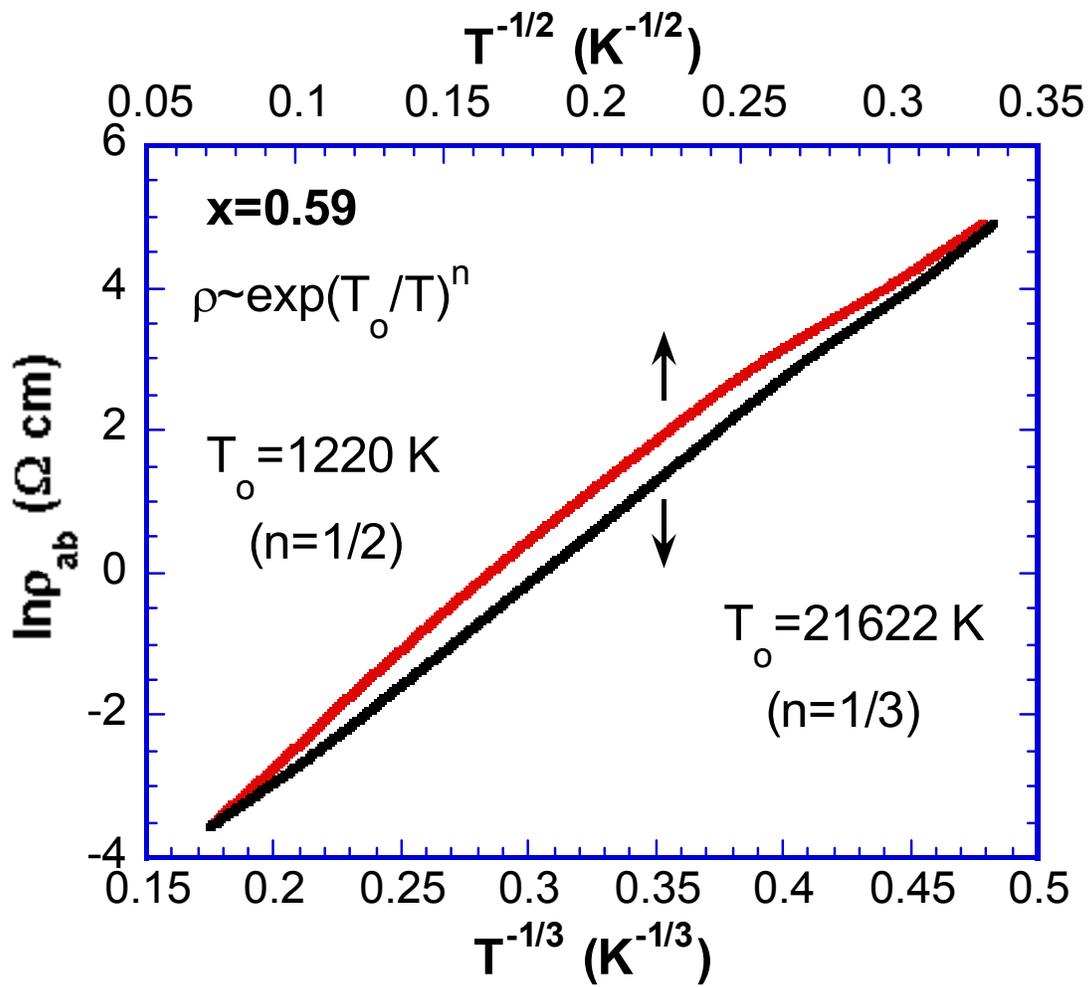

Fig.5

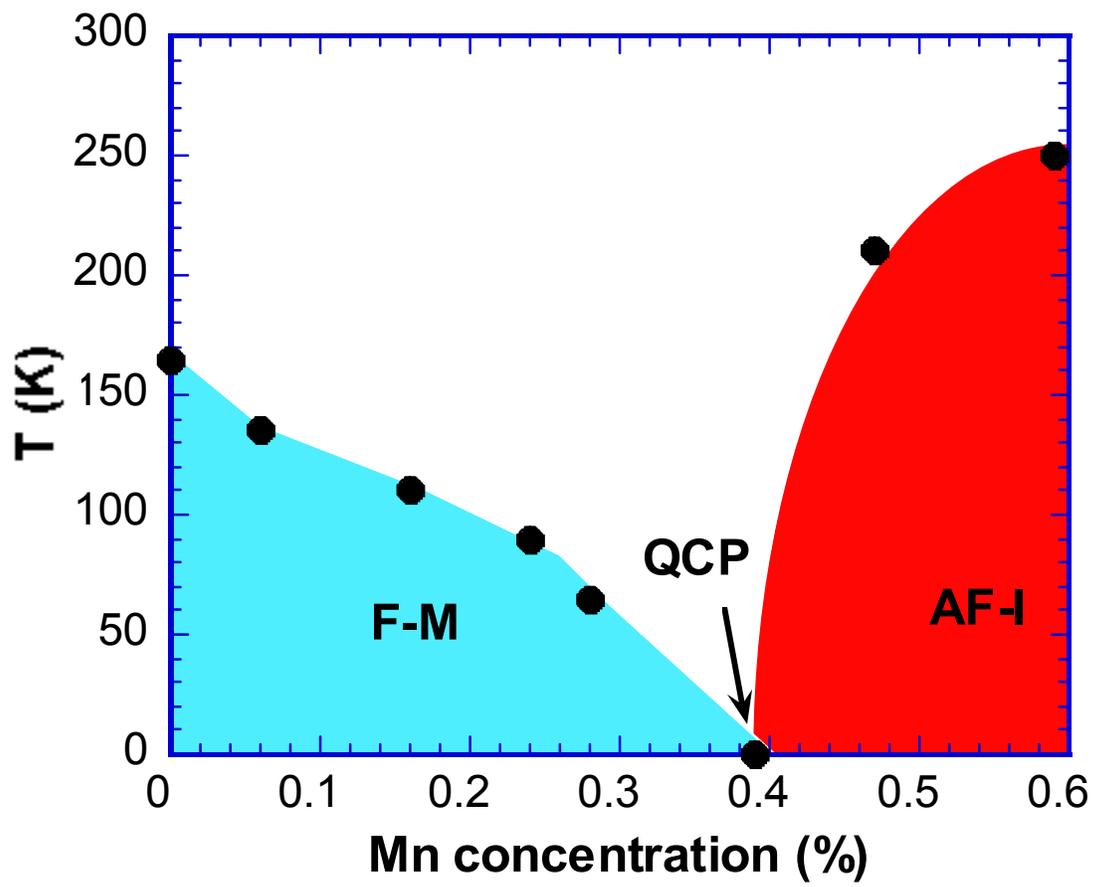

Fig.6